\documentclass{article}

\usepackage{arxiv}

\usepackage[utf8]{inputenc} 
\usepackage[T1]{fontenc}    
\usepackage{url}            
\usepackage{caption}
\usepackage{multirow}
\usepackage{dcolumn}
\usepackage{longtable}
\usepackage{threeparttable}
\usepackage{amsfonts}       
\usepackage{nicefrac}       
\usepackage{microtype}      
\usepackage{lipsum}
\usepackage{placeins}
\usepackage{graphicx}
\usepackage{rotating}
\graphicspath{ {./images/} }
\usepackage{natbib}
\setcitestyle{aysep={}}
\usepackage[hidelinks]{hyperref}

\title{The Need for A Recurring Large-Scale Benchmarking Survey to Continually Evaluate Sampling Methods and Administration Modes: Lessons from the 2022 Collaborative Midterm Survey\thanks{We thank Qi (Amelia) Li for outstanding research assistance to obtain population benchmark data and Grace Beals, Offir Ben-Salmon, Sohinee Bera, and Roxana Muenster for their excellent work on this project.}}

\author{
 Peter K. Enns \\
  Professor of Government and Public Policy, Cornell University\\
 Co-founder and Chief Data Scientist, Verasight\\
  \texttt{peterenns@cornell.edu} \\
   \And
 Colleen L. Barry \\
  Dean and Professor, Brooks School of Public Policy, Cornell University\\
  \texttt{cbarry@cornell.edu} \\
  \And
 James N. Druckman \\
  Professor, Department of Political Science, University of Rochester\\
  \texttt{jdruckma@ur.rochester.edu} \\
  \And
 Sergio Garcia-Rios \\
  Assistant Professor, Lyndon B. Johnson School of Public Affairs\\
  Associate Director for Research, Center for the Study of Race and Democracy, University of Texas, Austin\\
  \texttt{garcia.rios@cornell.edu} \\
 \And
 David C. Wilson \\
  Dean and Professor, Goldman School of Public Policy, University of California, Berkeley\\
  \texttt{dcwilson@berkeley.edu} \\
\And
 Jonathon P. Schuldt \\
  Associate Professor of Communication and Public Policy, Cornell University\\
  Executive Director, Roper Center for Public Opinion Research\\
  \texttt{jps56@cornell.edu} \\
}

\begin{document}
\maketitle

\thispagestyle{empty}
\setcounter{page}{0}
\setcounter{footnote}{0}
\clearpage

\begin{abstract}
As survey methods adapt to technological and societal changes, a growing body of research seeks to understand the tradeoffs associated with various sampling methods and administration modes. We show how the NSF-funded 2022 Collaborative Midterm Survey (CMS) can be used as a dynamic and transparent framework for evaluating which sampling approaches---or combination of approaches---are best suited for various research goals. The CMS is ideally suited for this purpose because it includes almost 20,000 respondents interviewed using two administration modes (phone and online) and data drawn from random digit dialing, random address-based sampling, a probability-based panel, two nonprobability panels, and two nonprobability marketplaces. The analysis considers three types of population benchmarks (election data, administrative records, and large government surveys) and focuses on national-level estimates as well as oversamples in three states (California, Florida, and Wisconsin). In addition to documenting how each of the survey strategies performed, we develop a strategy to assess how different combinations of sampling approaches compare to different population benchmarks in order to guide researchers combining sampling methods and sources. We conclude by providing specific recommendations to public opinion and election survey researchers and demonstrating how our approach could be applied to a large government survey conducted at regular intervals to provide ongoing guidance to researchers, government, businesses, and nonprofits regarding the most appropriate survey sampling and administration methods.    
\end{abstract}


\section{Introduction}
Surveys continue to be an important data source for government, media, political campaigns, and business, as well as public opinion, election, and other academic research. Yet, it is harder than ever to conduct surveys that meet the various needs of those conducting them. In an interview with \emph{Bloomberg Businessweek}, Former Director of the US Census Bureau, Robert Groves, went so far as to say, “[i]f we attempt to live through the 21st century with traditional survey methods producing our statistics alone, we’re in trouble” \citep{Pickert:2023}. According to the article, Groves’ grim assessment cited falling response rates as well as the “horrible cost-per-unit of gathering these data.”\footnote{In the same article, Erica Groshen, a former commissioner of the US Bureau of Labor Statistics, expressed similar concern about declining response rates and resulting implications on the representativeness and accuracy of government surveys, saying, “If this continues forever, if you just stretch these lines out, we’re going to be getting to a place where it’s hard to think that there won’t be some biases…I think any data user has to be always asking themselves, ‘How good are these data I’m basing these decisions on?’”}

Indeed, declining response rates, associated nonresponse bias, and the expense of longstanding probability-based methods represent some of the most salient challenges facing survey research. Nonresponse bias results when those who take the survey differ systematically from the underlying population and is particularly concerning when the nature of the bias is unknown or cannot be adjusted with poststratification weights. Concerns about nonresponse bias have increased in recent years as response rates have decreased. A recent political example comes from \citet{Cohn:2022b} who reports that in 2022, registered Democrats were 28 percent more likely than Republicans to respond to \emph{New York Times} Senate polls \citep[also see][]{Clinton:2021,Clinton:2022}. 

Declining response rates and the associated increase in costs of probability-based survey methods pose another challenge for survey research. \citet{Kennedy:2019} estimate that between 1997 and 2018, Random Digit Dial (RDD) telephone survey response rates declined from 36 percent to 6 percent and these declines have continued across survey modes \citep{Pickert:2023}. In 2022, just 0.4 percent of dials resulted in completed interviews in \emph{New York Times} surveys \citep{Cohn:2022a}. The cost increases associated with declining response rates are particularly salient for RDD and address-based sampling (ABS). Many researchers, media companies, and nonprofit organizations that depend on surveys often cannot afford traditional probability-based survey methods—even for studies of the general public.

Cost considerations are compounded by a growing recognition of the importance of studying variations in attitudes and behaviors across geographic regions to better understand election and social outcomes and political views, as well as considering subgroup heterogeneity and populations often underrepresented in traditional surveys (e.g., racial and ethnic minority groups, low-income individuals). Although survey research has a long and important history of studying minority and disadvantaged populations  \citep[e.g.,][]{DuBois:1899}, the sample size of most general population surveys limits subgroup analysis \citep{Welles:2014} often making other sampling strategies necessary \citep{Barreto:2018, Jackson:2004, King:2021}. 

In a 2008 symposium on internet surveys and election studies, \citet{Terhanian:2008} argued that multi-mode, multi-sampling-frame, multi-sampling-method studies could potentially enhance sample representativeness while reducing cost per interview. Consistent with this perspective, and in response to the challenges noted above, recent years have seen a major increase in the number of surveys that combine sampling methods (e.g., randomly selected addresses or phone numbers, convenience samples), administration modes (e.g., phone, online, in-person), or both. Developments in statistical modeling to weight data collected through nonprobability methods \citep[e.g.,][]{Lee:2006, Rivers:2007} or data from probability and nonprobability sources \citep[e.g.,][]{Fahimi2015} have also contributed to the rise of surveys that combine sampling approaches and/or administration modes. We have also seen new forms of survey data collection via the internet and new sample opportunities with internet panels and crowdsourcing.

Multiple administration modes further complicate survey research. Administration modes include in-person interviews, phone interviews, paper questionnaires, and online (web) surveys, but some of these modes can be divided even further. For example, phone surveys can take place via a live phone interview or interactive voice response (IVR); online surveys can be taken on a computer, a tablet, or a cell phone. Each of these administration modes (and sub-modes) can produce a different survey experience, including how respondents understand or interpret questions, the considerations that come to mind during the survey, and survey duration. Likewise, while sampling methods can be grouped into probability-based samples, nonprobability-based samples, and hybrid samples that combine probability and nonprobability samples, there are multiple types of probability and nonprobability samples. Probability sampling methods include strategies such as random digit dialing (RDD), random address-based sampling (ABS), and randomly selecting from a sampling frame, while nonprobability sampling methods can include approaches such as recruiting respondents through online advertisements, snowball samples, or through crowdsourcing tools like Amazon Mechanical Turk or sample aggregators/marketplaces like Cint (formerly Lucid). Online nonprobability samples have become ubiquitous and vary widely in quality, ranging from unweighted pure volunteer or convenience samples to sophisticated weighted quota-based samples.

In response to these changes in survey research, a growing literature has emerged to evaluate the various approaches. The results are mixed. Analyzing the accuracy of different sampling methods in estimating 40 benchmark variables, \citet{MacInnis:2018} found that probability samples yielded more accurate estimates than web samples that combined probability and nonprobability sampling, which in turn, outperformed web samples that used opt-in sampling exclusively. Additional support for the use of probability-based samples comes from a recent Pew Research Center report \citep{Mercer:2024}, which concludes that online opt-in (nonprobability) surveys can produce misleading results \citep[also see][]{Kennedy:2016}. However, a separate Pew Research Center benchmarking study found that for election polling nonprobability surveys were \emph{as accurate or more accurate} than probability surveys \citep[][7]{Mercer:2023}. Reaching a similar conclusion, \citet{Enns:2021} analyzed 355 surveys from the final two months of the 2020 presidential election and found that surveys that combined probability and nonprobability sampling were the most accurate, followed by nonprobability surveys, and then by probability-based samples \citep[also see][]{Collins:2021, Jackson:2015, Panagopoulos:2021, Vavreck:2008}.

This expanding literature has produced many valuable insights, but the variability in findings highlights a key lesson: Assessing various survey methods/modes is an ongoing challenge that will change depending on the research goal and over time as nonresponse patterns, demographics, technology, and costs change. Currently, there is no standard for evaluating the tradeoffs associated with employing various combinations of sampling methods and survey modes, or how these tradeoffs might vary across research questions and goals. Further, the ideal combination of administration and sampling modes is likely to continue to shift as social and technological conditions evolve.

To begin to address these challenges, we utilize data from the NSF-funded 2022 Collaborative Midterm Survey (CMS) \citep{Enns:2022b}, which interviewed 19,820 respondents using two administration modes (phone and online), four general population sampling methods (RDD, ABS, probability-based panel, and online nonprobability), and two samples of registered voters (ABS and SMS/text).\footnote{Online surveys could be taken by computer, tablet, or phone. While we do not analyze these separately, the survey data indicate device type and the archived methodological details include survey screenshots of all questions for all device types for all vendors. This level of detail offers a valuable opportunity for future research to assess potential differences by survey taking experience. The 2022 CMS was reviewed and approved by Cornell University IRB (IRB0144898).}  Further, the online nonprobability samples included two different nonprobability panels as well as two aggregators of nonprobability data, allowing further insight into potential variation across nonprobability sample sources. The CMS also asked numerous questions that allow direct comparison to population benchmarks, including standard comparisons, such as the 2022 Midterm election outcomes and questions asked on large government surveys like the American Community Survey and Current Population Study, as well as novel questions that can be compared directly to administrative data, such as marriage licenses and birth certificates. 

We show how this unique combination of multiple administration modes and sampling approaches with various population benchmarks offers a framework for evaluating which sampling approaches—--or which combinations of
approaches—--are best suited for various research goals. This  framework is designed to help evaluate tradeoffs related to “fit for purpose” \citep{Cornesse:2020,Groves:2005,Hillygus:2022} recognizing that researchers may have different goals when conducting a survey, such as being primarily interested in generating accurate point estimates of population values, testing a hypothesis about causality, maintaining the continuity of a time series, and so on. While we envision this framework being applied to a well-funded survey at regular intervals to allow ongoing assessment and benchmarking strategies for different survey samples and modes, we use the CMS illustrate how the framework would be applied. 

We first evaluate the accuracy of the overall CMS sample combining all methods. We then look at the findings for specific sampling approaches and modes. As a final step, we evaluate how different combinations of sampling approaches compare to different population benchmarks. While there is no single-best approach to survey research (optimal survey strategies will always depend on the research question), our goal is to show how these data can be used to inform researchers about tradeoffs of using different administration modes and sampling methods for various research goals and to offer a preliminary list of key lessons learned that can be immediately incorporated into election surveys and public opinion research.

While a number of previous studies have incorporated different sampling approaches and population benchmarks, our article differs from prior work in several ways. First, in contrast to many studies, our goal is not to identify which methods produce the most accurate results. Rather, we seek to highlight how different types of benchmarks and different analytic goals and considerations can lead to different conclusions. We make some specific recommendations related to election studies, but we also see value in helping identify the complexities and tradeoffs associated with different approaches. Second, ours is the first large-scale election study to evaluate multiple approaches within various categories (e.g., we consider multiple types of probability and non-probability sampling methods). Finally, we present a method for evaluating various \emph{combinations of methods}. We hope this approach is utilized by others to further evaluate strategies for combining sampling and administration approaches to balance sometimes competing goals of accuracy and cost.

The paper proceeds as follows. First, we provide a detailed overview of the CMS data and how other researchers can access the data for their own analyses (or to replicate our analysis). We then discuss the three population benchmarks used in our analysis (2022 Midterm Election outcome, number of children born in the last 10 years, and access to paid and unpaid internet) and report weighted and unweighted estimates for each administration mode and sampling approach as well as the overall sample for the adult population and the oversamples for the residents of the states of California, Wisconsin, and Florida. We also introduce a novel strategy for analyzing combinations of methods based on randomly subsampling particular samples within the overall CMS. While this strategy could be used to analyze any combination of administration modes or sample strategies (provided sample sizes are large enough) within the CMS—or any other survey that uses multiple methods—we follow a substantial amount of research in recent years \citep[e.g.,][]{Ansolabehere:2014, Enns:2021, Kennedy:2016, Jerit:2023, Mercer:2023} and focus this part of the analysis on probability and nonprobability samples. The paper concludes with a summary of lessons learned and a discussion of how the CMS’s multiple administration and sampling approaches combined with varied population benchmarks could be applied to a large ongoing government survey to offer a framework to help government, business, nonprofits, and academic researchers continually track the most appropriate survey methods for various research goals.

\section{Overview of 2022 CMS Data}
\label{sec:headings}
We rely on the 2022 Collaborative Midterm Survey (CMS) for our analysis \citep{Enns:2022b}. The CMS was conducted from October 26, 2022 to November 25, 2022 and includes 19,820 respondents. Samples represent the populations of English- and Spanish-speaking non-institutionalized U.S. adults and registered voters, including oversamples of California, Florida, and Wisconsin residents. As noted above, the CMS is ideal for our research purposes because it includes phone and online administration modes as well as multiple sampling approaches.

Figure 1 presents the sample sizes of the various sampling and administration modes in the 2022 CMS. All surveys were offered in English and Spanish. To keep the focus on the administration mode and sample approach and to ensure our analysis is not misinterpreted as a competition between the organizations or firms that collected the data, we do not link the names of organizations to data in the paper, but in the interest of full transparency this information can be found in Appendix Table A-1.\footnote{While we do not connect specific data providers with specific results in the main body of the paper, readers should be aware of the overall data sources. CMS data were collected through three data collection partners: SSRS, Gradient Metrics and Survey 160, and a team of researchers from the University of Iowa. Sample providers include SSRS, Gradient Metrics, Survey 160, YouGov, Dynata, Paradigm and their data partners, Ipsos and their partner data providers, and the Iowa Social Science Research Center. Response rates (probability samples) were 2.2\% SSRS (AAPOR RR3), 1.3 percent English-speaking phone sample (AAPOR RR4), 2.9\% Spanish-speaking phone sample (AAPOR RR4), 0.8\% SMS-to-web, 6\% mail-to-web (registered voters), and 4.3\% mail-to-web (adults) (AAPOR RR3) and completion rates (nonprobability samples) were 46.9\% combined Dynata, Paradigm and their partners, 36.7\% Ipsos and their partners, and 80.7\% YouGov.}  Moving from left to right in Figure 1, we see the CMS included three types of probability-based samples: a probability-based panel, random ABS (mail-to-web), and RDD (cell and landline). Although larger sample sizes for the ABS and RDD samples would have been preferable, these sample sizes correspond with 95 percent sampling margin of errors of approximately +/- 4.4 percent and 5.9 percent, respectively, which still allow for useful comparison.

The middle section of Figure 1 shows the CMS included two different nonprobability panels and two ``aggregators'' or ``marketplaces'' that combine multiple nonprobability sources. \citet{Enns:2022} have shown that it is increasingly common for survey firms to outsource data collection, often to data aggregators, who further outsource data collection. These samples offer an important assessment of this sampling strategy. The right section of Figure 2 shows two samples of registered voters. We only analyze these samples separately in the analysis of reported Midterm vote.

\begin{figure}[hbt!]
      \centering
      \includegraphics[width=1\linewidth]{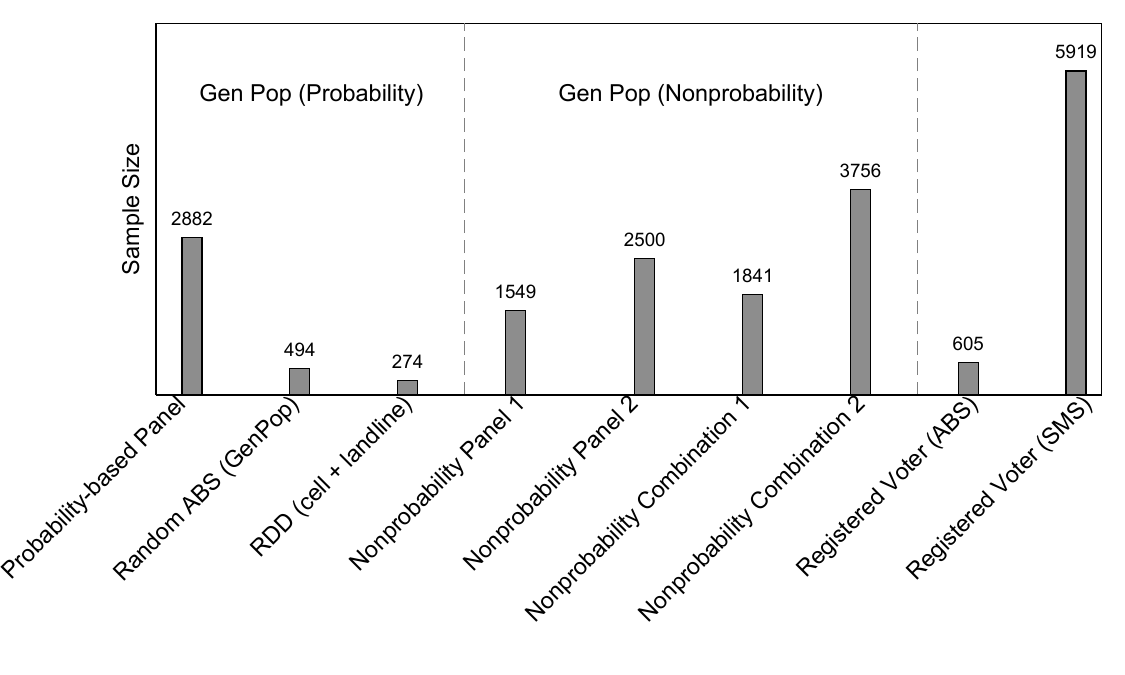}
      \caption{Sample Size by Administration Mode and Sampling Approach in the 2022 Collaborative Midterm Survey. 
  \textit{Note: Random ABS GenPop also includes respondents from the WI oversample. RDD is the only non-web-based administration mode.}}
      \label{fig:fig1}
  \end{figure}

While this is a single survey, we believe the CMS offers a transparent and dynamic framework that could be used for ongoing evaluation of which sampling methods and combinations of methods are best suited for various research purposes. In addition to evaluating combinations of sampling approaches and having a sample size larger enough to evaluate subgroups, the CMS developed a process to encourage methodological innovation and was able to quickly make data and all methodological information public. If repeated at scale, this combination of features could provide a road map to government, business, academic and policy surveys seeking to identify the ideal methodological approaches for their research needs. Below, we offer three examples from the CMS for how benchmarks from this type of survey can be used to inform methodological decisions. 

\FloatBarrier
\section{Analysis of Administration Mode and Sampling Approaches Using Three Types of Population Benchmarks}
\label{sec:headingsAdmin}
Our analysis uses benchmarks to evaluate the accuracy of various survey approaches and combinations of these approaches at the national level and in the three oversampled states of California, Florida, and Wisconsin. While accuracy is of interest, we do not view this as a “horse race” across approaches. Instead, our analysis is designed to assess the extent to which accuracy varies within sampling mode, as the analysis includes multiple sources of probability and non-probability methods as well as potential variability across benchmarks and geographic regions. Thus, we hope our analysis of these benchmarks helps researchers evaluate “fit for purpose” \citep{Cornesse:2020, Groves:2005, Hillygus:2022} and tradeoffs of various approaches as they might relate to specific research goals and considerations such as sample size, cost, and other goals. 

We focus on three types of benchmarks for our analysis.\footnote{The CMS includes standard demographic questions such as race, ethnicity, region, and gender, which can be compared to national population estimates from the decennial census. We do not use these demographics as benchmarks because we use them as weight variables. We believe it is crucial to incorporate weights into our analysis to accurately represent how the data would be used by a researcher or practitioner.}  First, we rely on an administrative political outcome: the Midterm Election. This is an ``administrative'' outcome because our population benchmark comes from government (administrative) election data. While the CMS includes questions on a range of substantive topics beyond elections, including attitudes toward democracy, authoritarianism, race, ethnicity, gender, economic views, and personal health, vote intentions prior to and reported vote after the election can be compared to the actual election outcome. We focus on the house election because unlike state ballot initiatives and senate and gubernatorial elections, which the CMS also asked about, all house districts had an election.\footnote{There were 23 U.S. House races without a Democratic candidate and 12 U.S. House races without a Republican candidate (\url{https://ballotpedia.org/U.S._House_elections_without_a_Democratic_or_Republican_candidate,_2022}).}  Second, we rely on a an administrative non-political outcome: childbirths. We use a question in the CMS that asked women in the sample (age 65 or younger) if they had a child in the previous 10 years (and if so, how many). We compare these responses to government administrative data on childbirths. It is extremely rare for surveys to include questions designed to compare directly to administrative records to estimate population values.\footnote{Most research uses government surveys, not administrative records, for population benchmarks, but see \citet{Kennedy:2016} for an exception that compared responses to a question about having a valid driver’s license with Federal Highway Administration data.}  Our third benchmark is government survey (non-political). This benchmark comes from the Census Bureau’s American Community Survey (ACS), a large government survey that is often used for population estimates \citep[e.g.,][]{Edwards:2014, McPhee:2022, Kennedy:2016}, in part because those selected to complete the ACS are required by law to do so.\footnote{\url{https://www.census.gov/programs-surveys/acs/about/acs-and-census.html}. Due to space constraints, we focus on these three benchmarks, but the CMS includes a question about marital status that can be benchmarked to government records on marriage licenses as well as additional questions that can be benchmarked to large government surveys; e.g., type of house/dwelling (ACS), social media use (CPS: Computer and Internet Supplement), retirement savings (CPS: Annual Social and Economic Supplement), employment status (CPS), and military service (numerous sources). We hope our research will encourage others to use our strategies and replication materials to offer further insights based on these benchmarks.} 

Our analysis relies on the publicly available version of the 2022 CMS archived at the Roper Center for Public Opinion Research \citep{Enns:2022b}. One of the advantages of the CMS is that the data include various survey weights for different research applications. For our purposes, however, we need to generate equivalent weights for each portion of the sample we wish to analyze. Generating identically constructed weights ensures that any observed differences are not a function of weighting strategy. We opt for a relatively straightforward approach, weighting by race/ethnicity, education, gender by age, and region (See Appendix 1 for full weighting details). 

\subsection{2022 Congressional House Vote}
Our first analysis considers the 2022 midterm vote. We focus on the house vote so that we can analyze responses from all states (all house districts included an election). Figure 2 presents the percent of respondents indicating a Republican vote intention (if surveyed prior to the election and they had not yet voted) or reported vote (if they voted early or were surveyed on or after Election Day) by each of the administration and sampling approaches presented above in Figure 1.\footnote{Following most election studies, we report percent indicating a Republican vote out of the two-party vote (i.e., those who indicated a vote intention or vote for either the Republican or Democratic candidate).}  The left panel presents results for the overall adult population and the right panel presents results for likely voters, defined as respondents who indicated they already voted, were certain to vote, or would probably vote and were very or extremely interested in the election (see Appendix 2 for exact question wording). Since the analysis focuses on vote choice, we also include the registered voter samples in the 2022 CMS, shown in the bottom two rows of Figure 2. 

For the overall adult population, although the actual result is outside the estimated uncertainty, the weighted CMS estimate (48.8 percent) is just 2.6 percentage points below the actual vote share (51.4 percent).\footnote{Cook Political Report 2022 National House Vote Tracker (\url{https://www.cookpolitical.com/charts/house-charts/national-house-vote-tracker/2022}).}  A nearly identical finding emerges for likely voters. This accuracy is reassuring considering our weighting approach was relatively simple, does not include any political variables (such as partisanship or past vote choice), and the four-week field period. To place this result in a broader context, we consider 538’s assessment of all pollsters that conducted at least five polls in the final 21 days before the election \citep{Rakich:2023}. While the 538 list and the CMS differ in sample size and survey field dates, the 538 list nonetheless offers a comprehensive overview of the accuracy of election polls. Among this list of 33 polling organizations, the 2.6 percentage point error would be the sixth most accurate, just 0.7 percentage points behind the most accurate and 1.8 percentage points more accurate than the median poll on the 538 list. The CMS’s overall accuracy highlights the potential benefits of combining multiple sampling approaches into a single survey. Of course, we are also interested in the results for each specific approach.

Looking to the second and third rows, weighted responses for “All Probability” match the election outcome almost perfectly for all adults and likely voters (51.2 percent and 51.4 percent, respectively) and are slightly more accurate than the corresponding results for “All Nonprobability” (48.8 percent and 48.2 percent). Turning to the individual survey approaches, the unweighted data underestimate the Republican house vote in every instance and fall outside of the estimated uncertainty seven out of nine times. In the weighted analysis, by contrast, the point estimates are closer to the election outcome in all but two approaches and the uncertainty intervals overlap the actual result in six out of nine instances. This pattern held for all adults and likely voters alike.

\begin{figure}[hbt!]
    \centering
    \includegraphics[width=.8\linewidth]{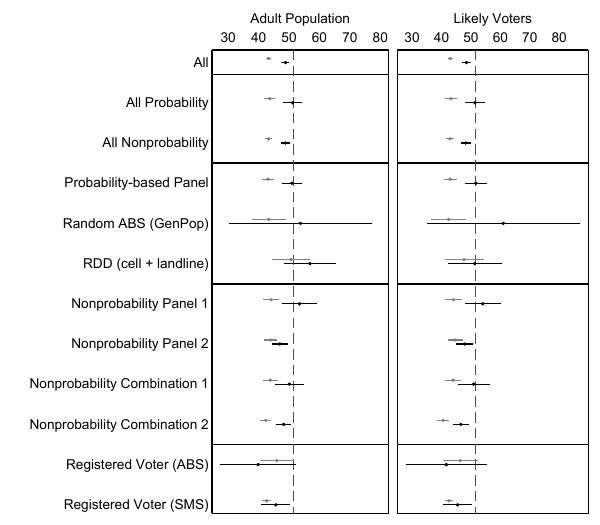}
    \caption{Congressional House Vote Intention/Reported Vote (percent Republican out of two-party vote) by 2022 CMS approach: Adults and Likely Voters.\\
  \textit{Note: Vertical red lines = actual Republican (two-party) vote (51.4\%). Horizontal lines depict 95\% uncertainty estimates. Gray corresponds with unweighted estimates and black corresponds with weighted estimates.}}
    \label{fig:fig2}
\end{figure}

The Registered Voter sample in the CMS is the least accurate, offering a cautionary note about assuming registered voter samples will necessarily offer a more accurate depiction of the electorate. While this is a surprise, the finding highlights the value of comparing the effects of different sampling frames while holding question wording, field period, and weighting strategy constant. Of course, with different elections or an alternate sampling strategy the findings for registered voters might be different.\footnote{To make sure this result was not a function of the weights we generated, we reanalyzed the registered voter samples in Figure 2 using the registered voter sample weights generated by the data vendor for the CMS. The conclusions remain the same. Using the vendor-generated  weights, the percent Republican vote (out of two-party vote share) is actually slightly less accurate than the weighting approach used in Figure 2 (42.2\% vs. 44.8\%).}  

\subsubsection{Evaluating Varying Combinations of Probability and Nonprobability Samples (House Vote)}
The above analyses considered each type of sample in the CMS individually. While it is common to focus on the accuracy of the different approaches in isolation, we are also interested in how various combinations of sampling approaches performed against the benchmarks. Here, we exploit the large CMS sample size of nearly 20,000 respondents and introduce a new strategy to evaluate the accuracy of varying combinations of probability and nonprobability samples. Although we could consider any combination of administration and sampling approaches in the CMS (e.g., combinations of ABS and online-probability samples, or combinations of nonprobability panels and nonprobability combinations/aggregators) with this particular analysis, we aim to build on the growing body of research that considers potential tradeoffs of probability and nonprobability samples more generally \citep[e.g.,][]{Ansolabehere:2014, Enns:2021, Kennedy:2016, Jerit:2023, Mercer:2023}.

The CMS includes 3,467 responses from probability-based samples of the general adult population (probability panel, RDD, and random ABS) and 9,646 from nonprobability-based samples of the general adult population (two nonprobability panels and two sources that combined data from multiple nonprobability vendors). Our goal is to evaluate how varying the proportion of the sample from these two sources affects the accuracy relative to our benchmarks. Currently, researchers who want to combine probability and nonprobability samples have very little guidance regarding how much of their sample should come from each source. To address this issue, we randomly sampled 1,000 respondents in the CMS data, varying the percentage that were drawn from probability and nonprobability portions of the CMS. We varied these samples of 1,000 CMS respondents from 0 percent probability (i.e., all 1,000 respondents were from the nonprobability portion of the CMS) to 100 percent probability (i.e., all 1,000 were from the probability portion) in increments of 10 percent. We repeated this procedure 500 times for each increment for a total of 5,500 random samples from the CMS data. Considering the similar findings observed for the adult and likely voter populations, we focus this part of the analysis on the adult population to maximize the number of potential respondents from which to sample.

The results are presented in Figure 3. The left (0) shows the results of 500 random samples of 1,000 CMS respondents; each sample being 0 percent probability-based (i.e., all 1,000 randomly sampled respondents were from the nonprobability-based component of the CMS). The vertical lines indicate the 95 percent uncertainty based on the 500 random samples. The next result reflects the estimate based on 500 random samples of CMS data where 10 percent (100 respondents) come from probability-based samples and 90 percent (900 respondents) come from nonprobability-based samples, and so on, all the way to the result at the right where 100 percent (1,000 respondents) come from probability-based samples and none come from nonprobability-based samples. These results show how the accuracy varies depending on the percent of the sample that is probability and nonprobability-based and associated uncertainty based on the variation across each of the 500 samples at each increment.

\begin{figure}[hbt!]
    \centering
    \includegraphics[width=.7\linewidth]{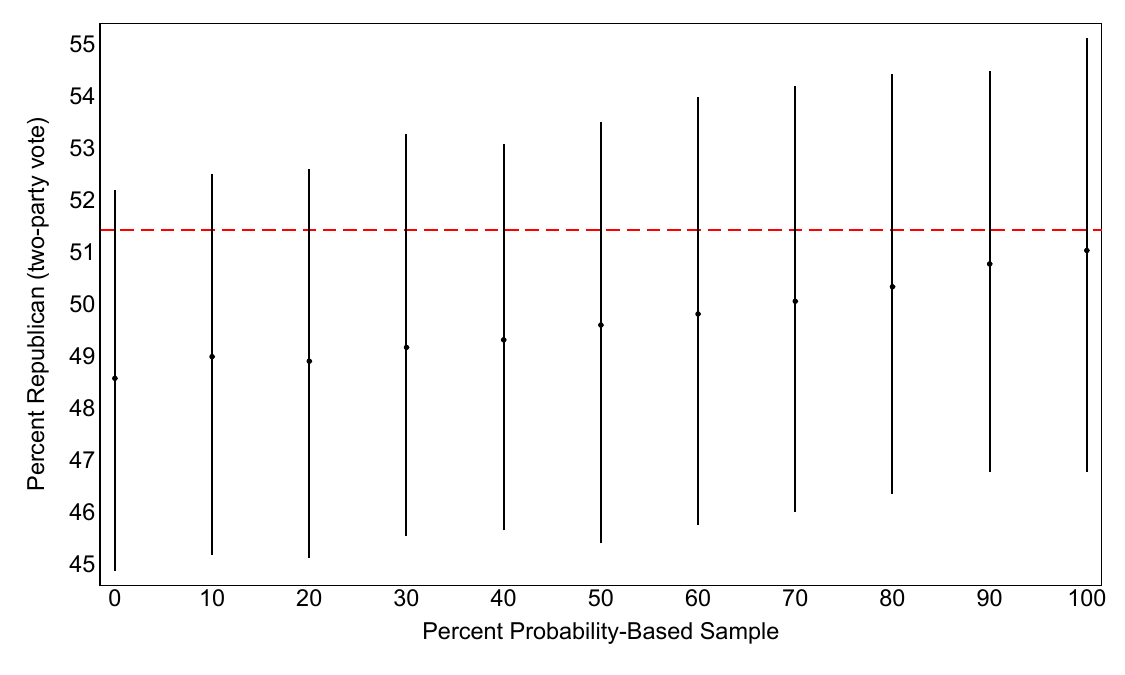}
    \caption{Estimated Republican House Vote (two-party vote share) ranging from 0\% probability-based sample (i.e., 100\% nonprobability) to 100\% probability-based sample (i.e., 0\% nonprobability), Adult Population.\\  
  \textit{Note: Each set of results along the x-axis based on randomly selecting 1,000 respondents 500 times.}}
    \label{fig:fig3}
\end{figure}

Two key findings emerge. First, in eleven out of eleven instances, the uncertainty interval overlaps the actual Republican two-party vote. Second, we see that increasing the proportion of probability-based sample yields a slightly more accurate estimate, such that a completely probability-based sample yields an estimate that is about two percentage points closer to the actual vote outcome than a completely nonprobability-based sample (though we cannot conclude that the estimates are statistically different). Thus, the results speak to the magnitude of the tradeoff that survey researchers face when deciding on whether to incorporate multiple sampling methods into their project, and at what proportions. When it comes to the 2022 House midterm election outcome, our analysis suggests only a small potential benefit to supplanting nonprobability-based sample with probability-based sample---a benefit that some researchers may conclude is not worth the financial cost or effort.

\FloatBarrier
\subsubsection{State-level findings (House Vote)}
Lastly, we turn to votes intentions/reported vote for the Republican and Democratic House candidates in California, Florida, and Wisconsin. As a reminder, with the exception of random address-based sampling in Wisconsin, our probability samples are not state-level probability samples, but rather are national probability-based samples, which we are using to conduct state-level analyses. A substantial amount of variability exists across sample approaches and across states. This variability highlights some of the challenges of state-level election surveys and shows that even if a particular approach (or combination of approaches) works well in one state, the approach may not work well in another state, even in the same election. Further complicating conclusions, even when a particular sampling method appears closer to a state's election outcome, differences across approaches are not statistically significant. 

\begin{figure}[h!]
    \centering
    \includegraphics[width=1\linewidth]{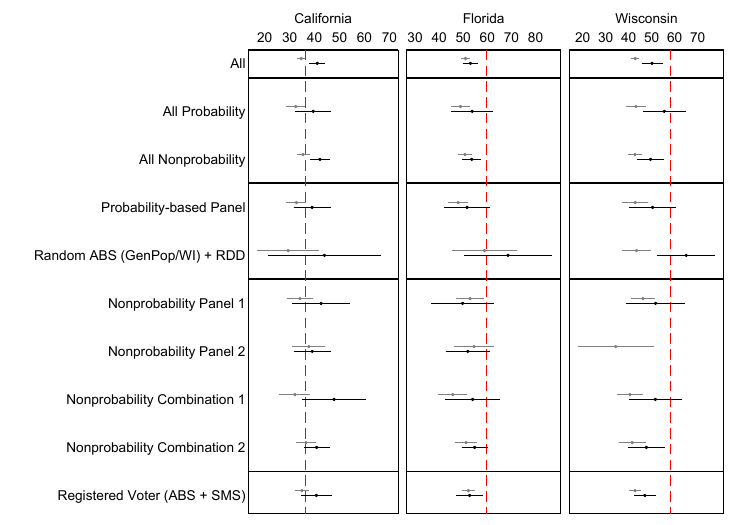}
    \caption{House Vote (Percent Republican, two-party vote share) in California, Florida, and Wisconsin.\\   
  \textit{Note: Vertical red lines show the actual house vote for each state. Horizontal lines depict 95\% uncertainty estimates. Some weighted estimates not shown due to insufficient observations for that sample type. Gray corresponds with unweighted estimates and black corresponds with weighted estimates.}}
    \label{fig:fig4}
\end{figure}

\FloatBarrier
\subsection{Administrative Benchmark: Childbirths}
Our second population benchmark considers a question that can be linked to administrative data: number of children born in the previous 10 years. The CMS asked women 65 or younger, “Have you given birth to any children in the last 10 years (i.e., the child or children were born in 2012 or later).”\footnote{Children born in most months of 2012 would have already turned 10 years old by the time of the survey. For this reason, we chose to specify 2012 in the survey question, ensuring all respondents anchor on the same start year, and that the start year allows respondents to use age of child (10 or under) as a heuristic for answering the question. Age of respondent was identified in the second question of the survey which asked what year respondents were born, and gender was identified by the third question which asked, “Do you describe yourself as a man, a woman, or in some other way?”}  Respondents who indicated yes were then asked, “How many children have you given birth to in the last 10 years?” The public version of the CMS data top-codes the number of births at 6, which includes those who indicated having more than 6 children in this period.

We use data on all birth records from the Centers for Disease Control and Prevention to identify the actual number of births in the United States since 2012.\footnote{\url{https://wonder.cdc.gov}} Because data on 2022 births are not yet available, we averaged the number of births in 2019, 2020, and 2021 to estimate 2022 births. The CDC presents birth data by age of mother. We subtracted births from women who would not have been eligible to take the survey. For example, births among those 16 and younger in 2021 would not be eligible to take the survey in 2022 and were not included in our total population estimate.

Figure 5 displays estimated childbirths by approach, relative to the actual number of childbirths in the U.S. since 2012 (42,091,245). We report unweighted (gray) and weighted (black) estimates to evaluate how the weighting process influences the conclusions (recall that the same weighting procedure was used for each approach). Overall, the weighted estimate based on the combined data (“All”) shows that the CMS estimate (42,195,773) was very close to the true population benchmark. As with the Midterm Election results above, the combination of sample sources in the CMS yield highly accurate results. However, the lower rows of the figure show somewhat different patterns than the previous analysis. In contrast to Midterm vote, the nonprobability-based samples were somewhat more accurate than the probability surveys overall, though certain probability modes (e.g., weighted RDD) were still among the most accurate. These results demonstrate that even within the same survey, different sampling approaches can lead to different patterns of accuracy. We believe this is an important and underappreciated finding in survey research.

\begin{figure}[h!]
    \centering
    \includegraphics[width=.8\linewidth]{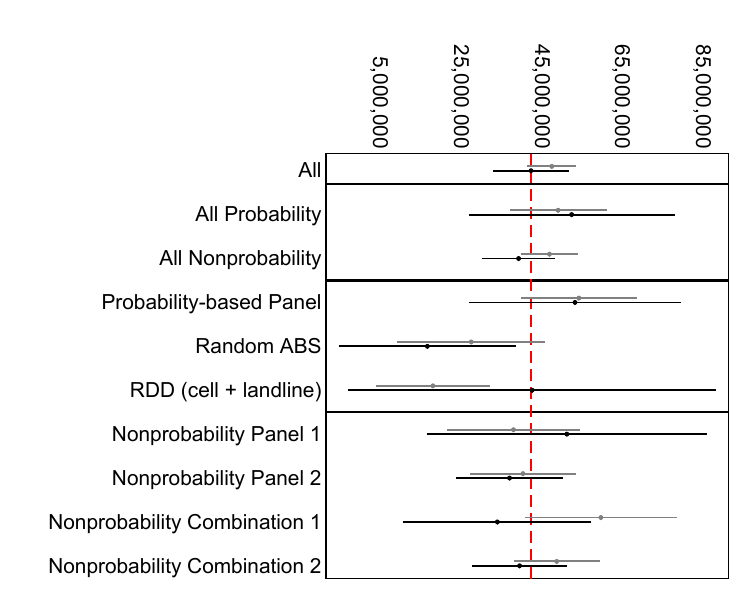}
    \caption{Number of Children born since 2012 by 2022 CMS sampling approach.\\   
  \textit{Note: Vertical red line shows the actual number of children born in the U.S. during this period (42,091,245). Horizontal lines depict 95\% uncertainty estimates. Gray corresponds with unweighted estimates and black corresponds with weighted estimates.}}
    \label{fig:fig5}
\end{figure}

\subsubsection{Evaluating Varying Combinations of Probability and Nonprobability Samples (Childbirths)}
Given potential interest in combining different types of samples within the same survey, we are again interested in how accuracy might vary with different combinations of probability and nonprobability samples. Figure 6 presents the results, once again based on 500 random draws of 1,000 respondents from the CMS dataset, at eleven different sample combinations (ranging from 0\% probability-based and 100\% nonprobability-based, to 100\% probability-based and 0\% nonprobability-based).

Some notable insights emerge. First, in all eleven cases, the 95 percent uncertainty interval overlaps the true population value. Second, this analysis reveals that the ideal proportion of probability to nonprobability sample is around 30 percent probability-based to 70 percent nonprobability-based, which is the combination level at which the point estimate (41,996,046) most closely approximates the true population value. From a fit for purpose perspective, this example highlights a couple of key takeaways for survey researchers interested in estimating the rate or incidence within the population. At least for childbirths, those interested in estimating the population value with a reasonable degree of uncertainty would be well served by any of the proportional combinations shown in Figure 6. By contrast, those seeking the most accurate estimate possible should choose a mix that is a predominately nonprobability-based, highlighting that pure probability-based samples do not always provide the most accurate insights.    

\begin{figure}[h!]
    \centering
    \includegraphics[width=.7\linewidth]{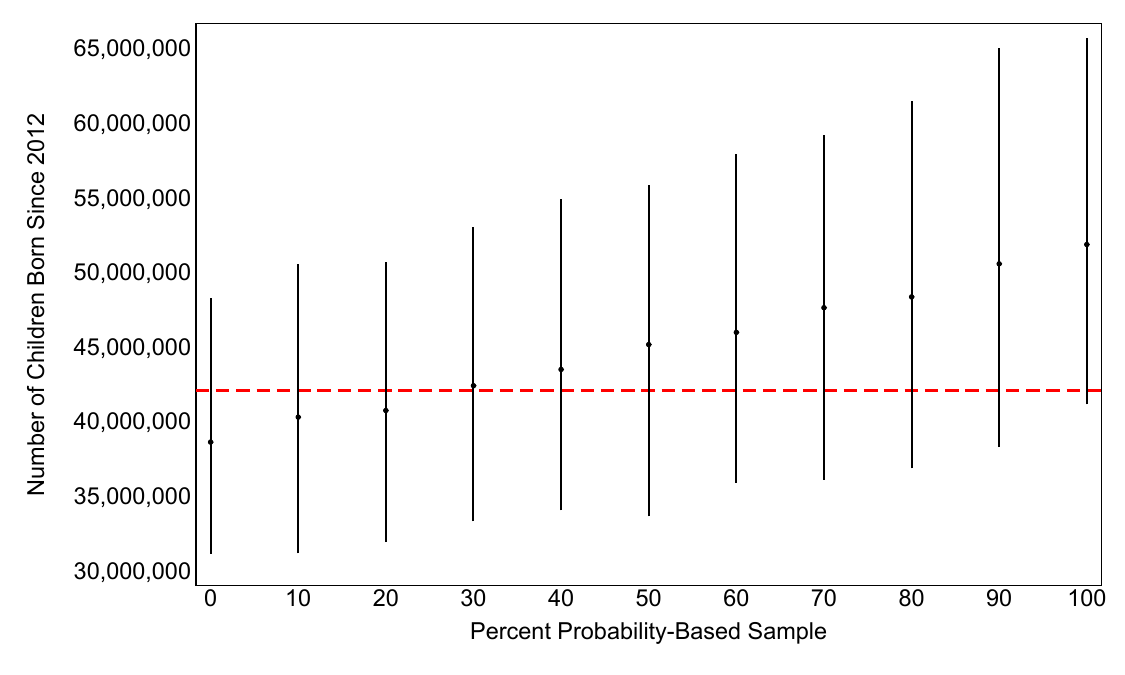}
    \caption{Number of Children born since 2012 by proportion of probability-based (vs. nonprobability-based) sample, each based on 500 random draws from the CMS dataset.\\  
  \textit{Note: Horizontal red line shows the actual number of children born in the U.S. during this period (42,091,245). Vertical lines depict 95\% uncertainty estimates.}}
    \label{fig:fig6}
\end{figure}

\subsubsection{State-level findings (Childbirths)}
Results from the oversamples in California, Florida, and Wisconsin are shown in Figure 7. Similar to the national-level findings (see Figure 5), overall, the combined CMS data yielded highly accurate estimates for the total number of children born in each of these three states since 2012. In contrast to the national-level findings where the combined (weighted) nonprobability-based samples yielded somewhat more accurate estimates, the combined (weighted) probability data surveys were the most accurate—although both sampling methods provided highly accurate estimates in all cases. Again, however, looking at the individual survey approaches reveals exceptions. 

\begin{figure}[h!]
    \centering
    \includegraphics[width=1\linewidth]{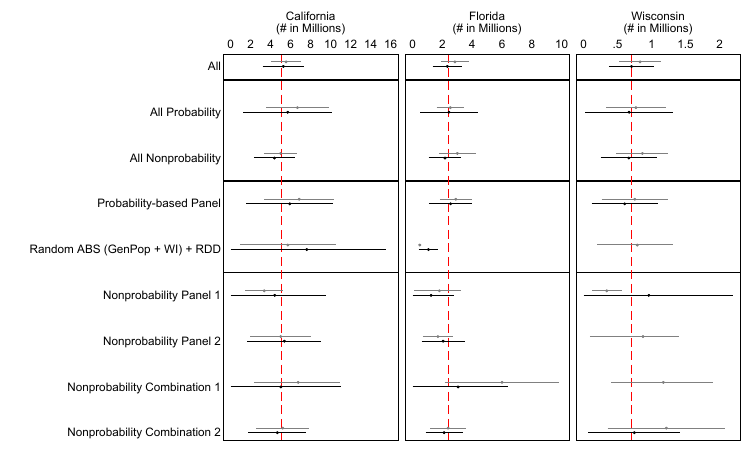}
    \caption{Number of Children Born Since 2012 in California, Florida, and Wisconsin.\\ 
  \textit{Note: Vertical red lines show the actual number of children born in each state during this period (i.e., CA: 5,121,438, FL: 2,401,935, and WI: 710,933). Horizontal lines depict 95\% uncertainty estimates. Gray corresponds with unweighted estimates and black corresponds with weighted estimates. Some weighted estimates not shown due to insufficient observations for that sample type.}}
    \label{fig:fig7}
\end{figure}

Perhaps most importantly, for Wisconsin, we see that for some approaches (Random ABS + RDD, Nonprobability panel 2, and Nonprobability panel 1) there are not enough observations to weight the data using the weighting procedure we selected. This shows how obtaining sufficient samples sizes within targeted geographic regions can be challenging for multiple sampling strategies. Although notable variation exists across survey approaches, in most cases the uncertainty around the estimates is too large to identify significant differences. Overall, comparing CMS data to a population benchmark from CDC administrative data, we see evidence that both probability and nonprobability samples yield accurate state-level estimates, suggesting that a critical first step is obtaining a sufficiently large sample size at the state level.

\FloatBarrier
\subsection{Survey-based Benchmark: Internet access (paid and unpaid)}
In this section we utilize a CMS question about respondents’ internet access that is worded identically to a question asked in the ACS (“At this house, apartment, or mobile home – do you or any member of this household have access to the Internet? Yes, by paying a cell phone company or Internet service provider. Yes, without paying a cell phone company or Internet service provider. No access to the Internet at this house, apartment, or mobile home”). Because 2022 ACS data are not yet available, we base population estimates below on the 2021 ACS. Looking at the left panel of Figure 8, the weighted CMS estimate for “All” respondents (89.5\%) underestimates paid internet access by about 3 percentage points. The unweighted estimate (91.9\%) is actually closer to the population benchmark. Importantly, the multi-mode design allows us to see that nonprobability online samples are the source of this pattern. Three out of four of these samples (bottom four rows) underestimate paid internet access and the weighted results consistently underestimate paid internet access by more. 

At first it might seem surprising that respondents taking the survey online are less likely to have paid internet. But the right-panel of Figure 8 helps explain this pattern. These same online respondents are more likely to have unpaid access to the internet. We speculate that these different patterns may reflect, in part, the different profiles of respondents who opt-in to complete online surveys, who may be more likely than respondents invited through random sampling procedures to obtain free access to the internet (e.g., through institutions such as schools, public libraries, community organizations, and businesses). This pattern, which we further explore in the analysis of combinations of approaches below, offers important implications for possible differences between probably and nonprobability internet samples.

\begin{figure}[hbt!]
    \centering
    \includegraphics[width=.75\linewidth]{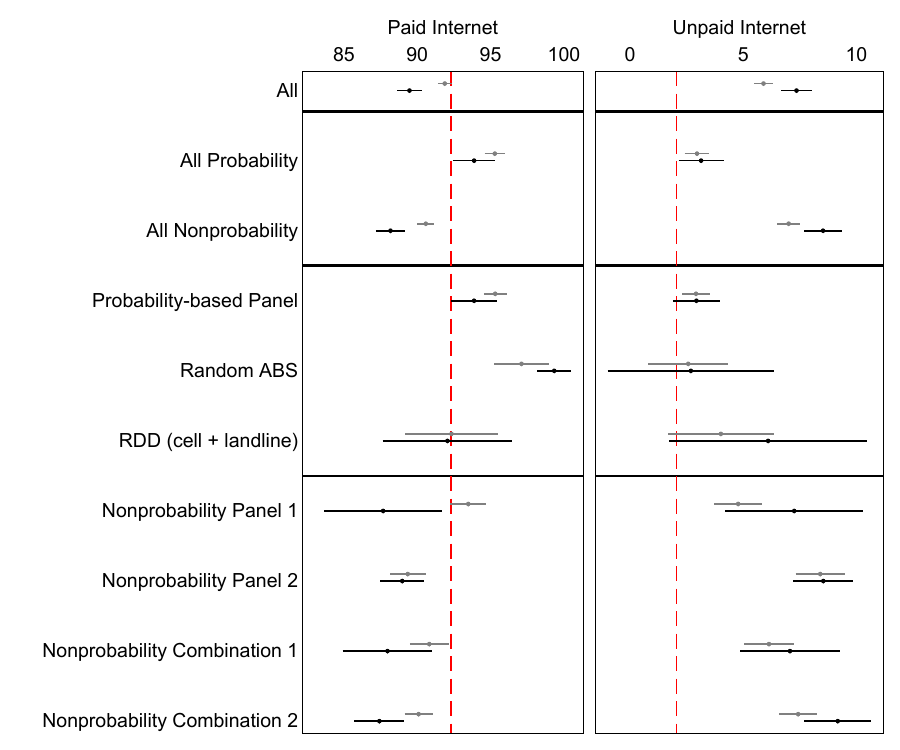}
    \caption{Paid and unpaid internet access.\\   
  \textit{Note: Vertical red lines = actual percent (92.3\% paid internet, 2.05\% unpaid). Horizontal lines depict 95\% uncertainty estimates. Gray corresponds with unweighted and black corresponds with weighted estimates.}}
    \label{fig:fig8}
\end{figure}

\FloatBarrier
\subsection{Evaluating Varying Combinations of Probability and Nonprobability Samples (Internet Acces)}
Figure 9 depicts estimated paid (top panel) and unpaid (bottom panel) internet access as a function of the percentage of probability-based (vs. nonprobability-based) sample, each based on 500 random draws of 1,000 respondents from the CMS dataset. The benchmark values (92.3\% for paid and 2\% for unpaid) are depicted by the horizontal red lines. While the estimated optimal combination differs some for paid and unpaid, considering both, 100 percent probability-based sample would yield the most accurate results. This finding stands in contrast to the results above focusing on childbirths, illustrating the importance of including multiple benchmark questions when trying to assess the accuracy of any survey. It is also possible, perhaps likely, that this result will change as internet access availability and ways of connecting to the internet change, illustrating the importance of not assuming that the accuracy of various administration and sampling modes will persist over time. Instead, it is crucial for the discipline to assess how administration and sampling modes perform in an ongoing manner. 

\begin{figure}[hbt!]
    \centering
    \includegraphics[width=.7\linewidth]{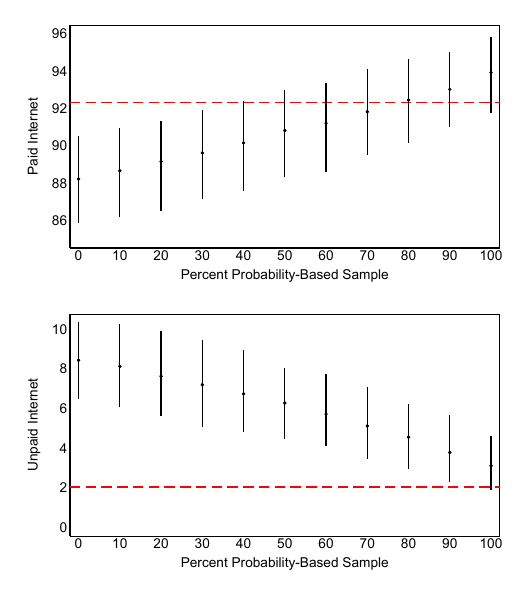}
    \caption{Paid (top panel) and Unpaid (bottom panel) internet access by proportion of probability-based (vs. nonprobability-based) sample, each based on 500 random draws from the CMS dataset.\\ 
  \textit{Note: Horizontal red lines show the estimated percent of unpaid and paid internet access among U.S. adults from the 2021 ACS. Vertical lines depict 95\% uncertainty estimates.}}
    \label{fig:fig9}
\end{figure}

\FloatBarrier
\subsection{State-level findings (Internet Acces)}
Turning to the state-level findings in California, Florida, and Wisconsin, the pattern of online nonprobability samples underestimating paid internet access and overestimating unpaid internet access continues, further suggesting that survey respondents recruited by online opt-in methods versus probability-based methods may have somewhat different profiles. From a fit for use perspective, researchers interested in fielding surveys related to internet technology attitudes and behaviors may wish to keep this difference in mind and will likely benefit by recruiting at least some portion of their sample through probability-based methods.   

We also see, however, more variability with the state estimates than we did with the national estimates across approaches. For example, looking at paid internet access, “All Probability” produces the most accurate estimate in California and “All Nonprobability” produces the most accurate estimate in Wisconsin. Similar variations emerge when we look at sampling approaches within probability and nonprobability. This variability further reinforces the challenges associated with surveys of particular geographic regions, as even when methods and sample sizes are similar across states, accuracy of the survey can vary. 

\begin{figure}[hbt!]
    \centering
    \includegraphics[width=1\linewidth]{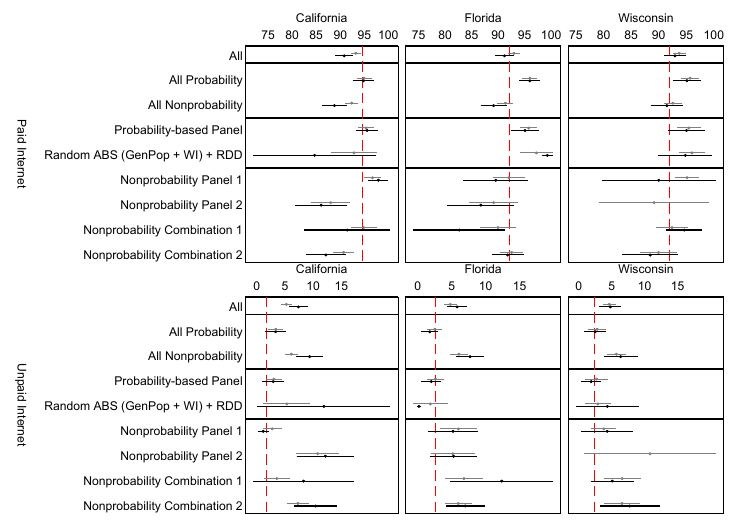}
    \caption{Paid (top panel) and unpaid (bottom panel) internet access in California, Florida, and Wisconsin.\\ 
  \textit{Note: Vertical red lines show the estimated percent of paid and unpaid internet access among state residents from the 2021 ACS. Horizontal lines depict 95\% uncertainty estimates. Gray corresponds with unweighted estimates and black corresponds with weighted estimates. Some weighted estimates not shown due to insufficient observations for that sample type.}}
    \label{fig:fig10}
\end{figure}

\FloatBarrier
\section{Implications for Election Surveys and Public Opinion Research}
The proliferation of surveys that combine different sampling methods and/or administration modes holds much promise, but also poses a challenge as researchers and practitioners seek to identify which approaches and combinations of approaches are best suited for various goals. To help navigate this process, we used the 2022 CMS to evaluate multiple approaches and their combinations. In this final section, we discuss some lessons learned that apply to election surveys as well as insights for the broader public opinion and survey research community. 

\subsection{Lessons for Election Surveys}
The above analysis of CMS data offers several insights for future election surveys. First, registration-based sampling does not necessarily outperform other sampling approaches when the goal is to understand house vote intentions and vote choice. Although the CMS overall and particularly its probability-based portion were highly accurate in estimating Republican vote share, the registered voter samples (ABS and SMS) did not perform as might be expected. It may be that these results reflect an invitation mode effect, as the registered voter sample sent survey invitations via mail and text. Alternatively, it may be that states that tend to lean Republican publicly release updates to their voter registration files less frequently, on average, than more Democratic states, which could introduce a bias to the sampling frame. Importantly, the observed patterns may vary for different types of elections. We hope future research investigates these, and other possibilities, but for now it appears that a registration-based sampling frame is not necessarily the most representative of vote outcomes.  

A second lesson relates to the RDD sampling. Although the CMS planned to interview a larger sample of RDD respondents, just 274 RDD respondents were interviewed. This unexpectedly low number highlights the need for projects using phone interviews to gather evidence of phone response rate capabilities, as recruiting respondents willing to take surveys over the phone with a live interviewer may grow increasingly challenging in years to come. Our analysis also offers lessons for election surveys that aim to reach specific populations of interest. Given the election-focus of the CMS, the subgroups of primary interest were adults in California, Florida, and Wisconsin. ABS samples combined with RDD (which was necessary given the small sample sizes of each) were relatively accurate across benchmarks. Indeed, the most accurate results came from the ABS and RDD sample with 112 respondents from California, 108 from Florida, and 275 from Wisconsin. These results show that probability-based sampling methods can beneft election surveys, even with relatively small sample sizes. However, this result stands in contrast to work on the 2020 presidential election \citep[e.g.,][]{Collins:2021, Enns:2021}, suggesting that this type of analysis should be conducted at regular intervals to help establish whether particular patterns are context specific or might generalize.   

\subsection{Lessons for Surveys Generally}
The overall accuracy of the CMS across election-related and non-election related benchmarks suggests that combining multiple probability and nonprobability methods may be a promising approach for survey research more broadly. Even when we do not know in advance which combination is ideal, overall, combining approaches seems to produce more accurate results than utilizing a single approach. 

A second and equally important insight is that the ideal combination of approaches varies across goals and can be counterintuitive. As a notable example, we found that online surveys over-estimate paid internet access. For this reason, when studying something like internet access that might be correlated with administration mode, traditional probability-based sampling may be expected to yield more accurate results. 

\subsection{A Large-Scale Benchmarking Survey to Continually Evaluate Sampling Methods and Administration Modes}
In closing, we hope that our analysis of the 2022 CMS data can provide a new model for identifying the advantages and disadvantages of various combinations of survey approaches. We envision this framework being applied to a large-scale survey at regular intervals to allow ongoing assessment and benchmarking strategies for distinct sampling strategies and multi-mode surveys. Researchers are often stuck wondering whether lessons learned from past surveys or surveys with other research goals generalize. We believe a large, regularly administered government-funded survey could build on the present work to provide up-to-date insights that help answer these questions. Further, although a survey of this magnitude would be expensive, it would \emph{save money} by allowing more accurate and cost-effective surveys among researchers, government, business, nonprofits, and media. 

The proposed framework is designed to help evaluate tradeoffs related to “fit for purpose” \citep{Cornesse:2020, Groves:2005, Hillygus:2022}, recognizing that researchers may have different goals when conducting a survey, such as being primarily interested in generating accurate point estimates of population values, testing a hypothesis about causality, maintaining the continuity of a time series, and so on. Those interested in generating highly accurate population estimates may be especially interested in the ideal proportion of probability to nonprobability sample, which may matter less to researchers interested in causal inference; meanwhile, someone interested in time series continuity will be interested in the relative accuracy of the methods used in previous surveys, and whether they continue to reach the same respondent groups as before.

Perhaps most importantly, and consistent with the need for an ongoing large-scale benchmarking survey, we found evidence that a single sampling approach does not outperform others. Despite the scientific foundations, probability-based surveys are not always the most accurate  (also see Enns and Rothschild 2021; Jackson 2015; Mercer and Lau 2023, 7; Panagopoulos 2021). In a perfect world, with no non-response bias, this would not be the case. But all samples have bias.  What sample (or combination of samples) is most appropriate depends on the nature of that sample, the mode, the topic being studied, and the purpose. Further complicating things, as technology and society continue to change, how samples and modes perform will continue to change. The discipline needs a recurring benchmarking survey that can provide researchers with ongoing information about how different sampling approaches perform for different topics and research goals.

While clear recommendations emerge from the above analyses, a core goal has been to show how the CMS can be used by researchers interested in evaluating methodological tradeoffs that are regularly faced in election surveys and public opinion research. Because the CMS data are publicly available, students, researchers, and practitioners can continue to build on the insights presented here. This work can evaluate other benchmarks and sample or mode combinations as well as examine the myriad of substantive questions in the CMS, such as those related to the economy, democracy and authoritarianism, political attitudes, and pro-social behavior. We hope these considerations and findings will offer a useful guide to public opinion and survey researchers going forward.

\bibliographystyle{chicago}  
\bibliography{references}  





\clearpage
\renewcommand{\thepage}{A-\arabic{page}}\setcounter{page}{1}
\renewcommand{\thesection}{Appendix \arabic{section}}\setcounter{section}{0}
\renewcommand{\thefigure}{A-\arabic{figure}}\setcounter{figure}{0}
\renewcommand{\thetable}{A-\arabic{table}}\setcounter{table}{0}

\centerline{\LARGE{Appendix}}

\section*{Appendix 1: Weighting Details}

We use ipfraking in Stata \citep{Kolenikov:2019} to generate the weights. For national-level estimates, race/ethnicity is coded as White non-Hispanic, Black non-Hispanic, Hispanic, other non-Hispanic, Education is coded as less than high school degree, high school degree, some college, four-year college degree, more than four-year degree. Age by gender coded as male 18-24, female 18-24, male 25-34, female 25-34, male 35-44, female 35-44, male 45-54, female 45-54, male 55-64, female 55-64, male 65+, female 65+. Region was coded as the four census regions (Northeast, South, Midwest, West) plus California, Florida, and Wisconsin. This ensures that respondents from the California, Florida, and Wisconsin oversamples were weighted to reflect the proportion of California, Florida, and Wisconsin residents in the population. For state-level estimates we use six gender by age categories instead of 12 and education is collapsed into four categories instead of five.

\section*{Appendix 2: Midterm Election Analysis Question Wording}

Exact question wording for the childbirth and internet analyses were included in the main text. The following questions were used to identify likely voters and vote intention/vote choice.

How likely are you to vote in [the November 8/today’s] election for Congress? Will not vote;
Less than 50-50 chance; Chances are 50-50; Will probably vote; Absolutely certain to vote;
Already voted

Which of the following statements best describes you? I did not vote in the election this November. I thought about voting this time – but didn't. I usually vote, but didn't this time. I attempted to vote but did not or could not. I definitely voted in the November 2022 Election.

How interested [are you/were you] in the election taking place [on November 8th / today / that
took place on November 8th] in (INSERT STATE FROM Q15A OR Q1 IF Q15A NOT
ANSWERED)? Extremely interested; Very interested; Somewhat interested; Only a little
Interested; Not at all interested 

Thinking about the election for US House of Representatives [that's being/that was] held on November 8, for whom [will you/did you/would you] vote [if you had voted] in your congressional district? The Democratic candidate; The Republican candidate; Other candidate

\section*{Appendix 3: Sample Size by Method and Survey Organization}

Figure 1 in the text reports sample size by administrative mode and sampling approach. Table A-1 provides two pieces of additional information. First, Table A-1 indicates the survey organizations and the specific data providers they used. In the main text, we omitted these names to keep the focus on the administrative mode and sample approach and to avoid any (incorrect) perception that this paper is somehow a competition between organizations or firms. The Collaborative Midterm Survey was indeed a collaboration with the three teams from SSRS, Gradient/Survey 160, and the University of Iowa working together on every step of the process from basic questionnaire design, to survey translations, to weighting, to methodological documentation. Table A-1 also provides sample size by California, Florida, and Wisconsin, the three states that were oversampled.
	
\begin{sidewaystable} 
\centering
\begin{threeparttable} 
\caption{CMS Sample Size by Method and Survey Organization} \label{tab:modes}
\begin{tabular}{lcccccc} 
\hline
                &                &                 & \multicolumn{3}{c}{State Oversample} & Total \\
Organization    &   \multicolumn{2}{c}{Method}    & CA & FL & WI & N= \\
\hline
SSRS                & Prob    & SSRS Probability Panel         &623 &615 &331 & 2,882 \\
SSRS                & Prob    & Mail-to-Web (Random ABS)       &0 &0 &183 & 183 \\
SSRS                & Nonprob &  Dynata                        &364 &301 &392 & 1,549    \\
SSRS                & Nonprob & Paradigm + partners$^\dag$     &253 &309 &303 & 1,841 \\
\hline
Gradient/Survey 160 & Prob    & SMS-to-Web (Registered Voters) &1,501 &1,580 &1,478 & 5,919 \\
Gradient/Survey 160 & Prob    & Mail-to-Web (Registered Voters)&150 &163 &159 & 605   \\
Gradient/Survey 160 & Prob    & Mail-to-Web (General Public)   & 86 & 80 & 77 & 311   \\
\hline
Univ. of Iowa Team  & Nonprob &  Ipsos + partners$^\dag$       & 644& 536& 319&  3,756 \\
Univ. of Iowa Team  & Nonprob &  YouGov                        & 250& 175& 37 & 2,500  \\
Univ. of Iowa Team  & Prob    & Iowa SSRC$^\ddag$ (cell \& landline)& 26& 28& 15 & 274  \\
\hline
\emph{Total N=}       &         &                          &   & & & 19,820 \\
\hline
\end{tabular} 
\begin{tablenotes}[para] 
$^\dag$ Paradigm recruited respondents through Cint, Pure Spectrum, MarketCube, Disqo, and Precision. Ipsos also recruited respondents through external sources but details about their data partners are not available.\\ $^\ddag$ Iowa SSRC = Iowa Social Science Research Center\\
Prob refers to probability-based sampling method and nonprob refers to nonprobability sampling. Total includes all states plus Washington DC. (Table reproduced from \cite{Enns:2023})
\end{tablenotes} 
\end{threeparttable} 
\end{sidewaystable} 

\end{document}